\documentclass[aps,prl,twocolumn,amsmath,amssymb,longbibliography,superscriptaddress,groupedaddress]{revtex4-2}
\usepackage{multirow}
\usepackage[separate-uncertainty = true,multi-part-units=single]{siunitx}
\usepackage{graphicx}
\usepackage[english]{babel}
\usepackage[utf8]{inputenc} 
\usepackage{ctable} 
\usepackage{booktabs}
\usepackage{float}
\usepackage{wasysym}
\usepackage{xcolor}
\usepackage{color, colortbl}
\usepackage{german}
\usepackage[colorlinks=true,allcolors=blue]{hyperref}
\usepackage{doipubmed}

\newcommand{\Kr}[1]{$^{#1}$Kr}

\definecolor{Lightblue}{rgb}{0.85,0.85,0.95}
\definecolor{Lightred}{rgb}{0.95,0.9,0.9}

\begin{document}

\title{Optical Excitation and Trapping of \Kr{81} }
\author{J. S. Wang}

\author{F. Ritterbusch}
\email{florian@ustc.edu.cn}

\author{X.-Z. Dong}
\author{C. Gao}
\author{H. Li}
\author{W. Jiang}
\author{S.-Y. Liu}
\author{Z.-T. Lu}
\email{ztlu@ustc.edu.cn }
\author{W.-H. Wang}
\author{G.-M. Yang}
\author{Y.-S. Zhang}
\author{Z.-Y. Zhang}

\affiliation{\medskip Hefei National Laboratory for Physical Sciences at the Microscale, CAS Center for Excellence in Quantum Information and Quantum Physics, University of Science and Technology of China, 96 Jinzhai Road, Hefei 230026, China}

\begin{abstract}
We have realized optical excitation, trapping and detection of the radioisotope \Kr{81} with an isotopic abundance of 0.9 ppt. The \SI{124}{nm} light needed for the production of metastable atoms is generated by a resonant discharge lamp. Photon transport through the optically thick krypton gas inside the lamp is simulated and optimized to enhance both brightness and resonance. We achieve a state-of-the-art \Kr{81} loading rate of \SI{1800}{atoms/h}, which can be further scaled up by adding more lamps. The all-optical approach overcomes the limitations on precision and sample size of radiokrypton dating, enabling new applications in the earth sciences, particularly for dating of polar ice cores.
\end{abstract}
\maketitle


\textit{Introduction.--- }\Kr{81} ($t_{1/2}=\SI{229}{ka}$) is the ideal isotope for radiometric dating of water and ice in the age range from thirty thousand years to over one million years \cite{Loosli1969, Lu2014}. Atom Trap Trace Analysis (ATTA) \cite{Chen1999}, an analytical method that detects single atoms via their fluorescence in a magneto-optical trap (MOT), has overcome the difficulties of detecting this isotope at the extremely low isotopic abundance levels of $10^{-14} - 10^{-12}$ in the environment, enabling a wide range of applications in the earth sciences \cite{Jiang2012, Lu2014}. The required sample size for \Kr{81} analysis is \SI{1}{ \mu L} (at standard temperature and pressure) of krypton, which can be extracted from \SI{10}{-}\hspace{0.02cm}\SI{20}{kg} of water or ice \cite{Tian2019,Jiang2019}.
\begin{figure}[b]
  \centering
  \noindent  \includegraphics[width=8.5cm]{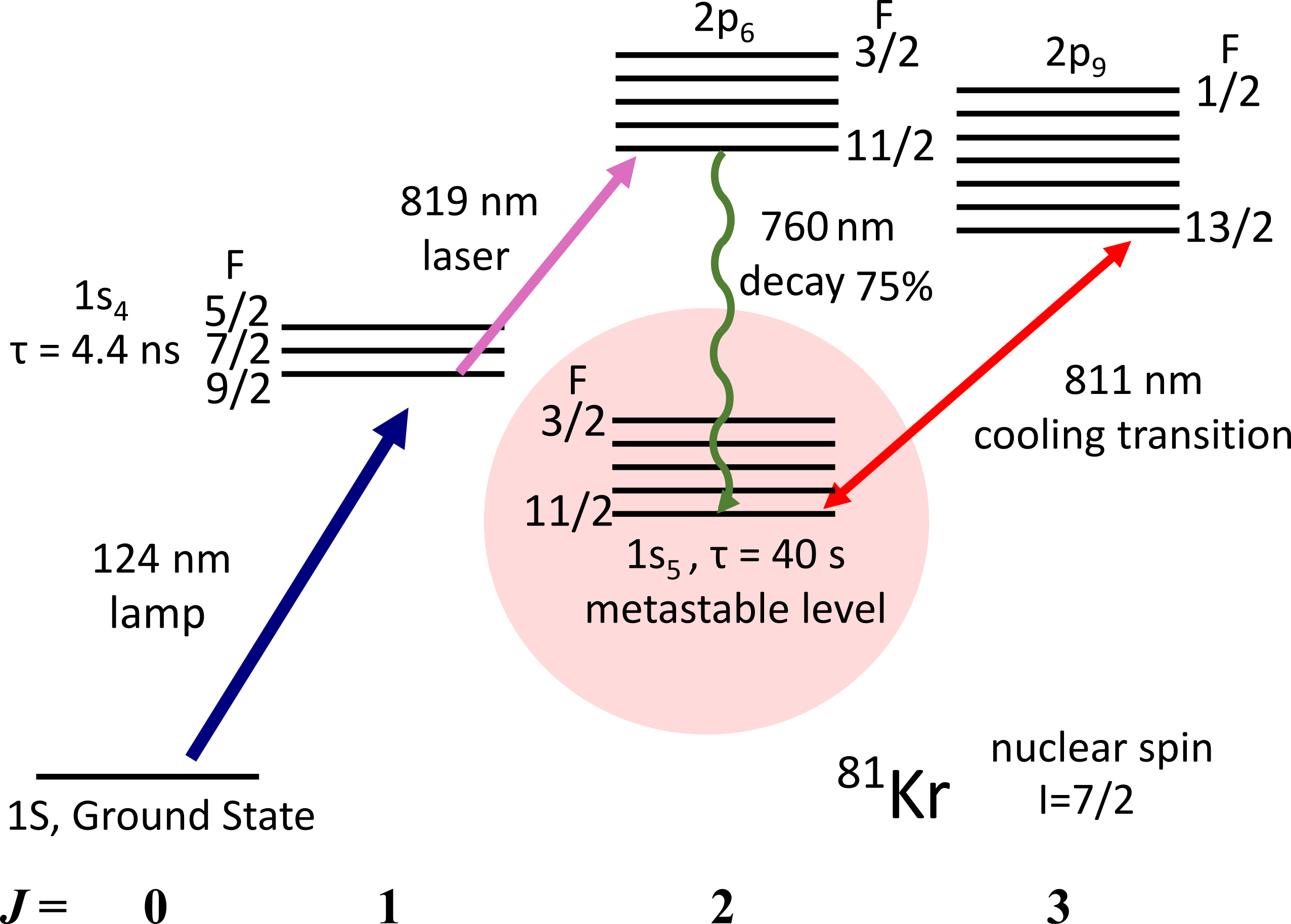}
  \caption {Optical excitation schematic for \Kr{81}.}
  \label{fig:energy}
\end{figure} The upper age reach is 1.3 million years at which the \Kr{81} isotopic abundance drops to $\sim$\hspace{0.05cm}\SI{2}{\%} of the modern level \cite{Jiang2019}. However, the sample size requirement is still too large and the upper age reach still does not go back far enough for \Kr{81} to unveil its full potential, especially for major questions in paleoclimatology such as the Mid-Pleistocene Transition \cite{Clark2006, Severinghaus2010}, the Greenland ice sheet stability \cite{Schaefer2016, Yau2016}, or the stability of ice caps on the Tibetan Plateau \cite{Thompson1821, Thompson2005, Hou2018}.\\
\indent Laser cooling and trapping of krypton atoms in the ground-level is not feasible due to the lack of suitable lasers at the required VUV wavelength. Instead, the krypton atoms first need to be transferred to the metastable level $1\text s_5$ where the $1\text s_5$\hspace{0.05cm}$-$\hspace{0.05cm}$2\text p_9$ cycling transition at \SI{811}{nm} can then be employed (Fig.\hspace{0.1cm}\ref{fig:energy}). The level $1\text s_5$ is $\sim$\hspace{0.05cm}\SI{10}{eV} above the ground level and, in operational ATTA instruments, is populated by electron-impact excitation in a discharge. However, the electron-atom collisions also destroy metastable atoms via de-excitation and ionization \cite{Hyman1978, Hyman1979, Mityureva2020}, leading to a metastable population fraction of only $10^{-4}$$-$$10^{-3}$. Meanwhile, krypton ions are implanted into the surfaces of the vacuum system leading to sample loss, and previously embedded krypton is sputtered out causing cross-sample contamination. These are currently the main limitations for reducing the sample size as well as extending the age range of \Kr{81} dating.\\ 
\indent These limitations of the discharge excitation can be overcome by optically exciting the krypton atoms from the ground to the metastable level. One realization is via off-resonant two-photon excitation to the level $2\text p_6$, requiring a laser at \SI{215}{nm}, followed by a spontaneous decay to $1\text s_5$. This mechanism has recently been demonstrated in a spectroscopy cell using a pulsed optical parametric oscillator laser \cite{Dakka2018}. Metastable production efficiencies up to \SI{2}{\%} per pulse (\SI{2}{mJ} per \SI{5}{ns} long pulses, \SI{10}{Hz} repetition rate) were achieved in the focus region of \SI{0.1}{mm} across. For this scheme to be applicable in a practical ATTA system, laser powers would have to be higher by several orders of magnitude. Another way to optically excite the krypton atoms is by resonant two-photon excitation via the \SI{124}{nm} and \SI{819}{nm} transition (Fig.\hspace{0.1cm}\ref{fig:energy}). Light at \SI{124}{nm} can be generated by four-wave-mixing in a mercury cell \cite{Albert2013,Wang2017} or by a free electron laser \cite{Chang2019}. At least \SI{1}{mW} of time-averaged power within the resonant linewidth of $\sim$\hspace{0.05cm}\SI{3}{GHz} would be needed for this excitation scheme.\\ \\
\begin{figure}[t]
  \centering
  \noindent  \includegraphics[width=8.5cm]{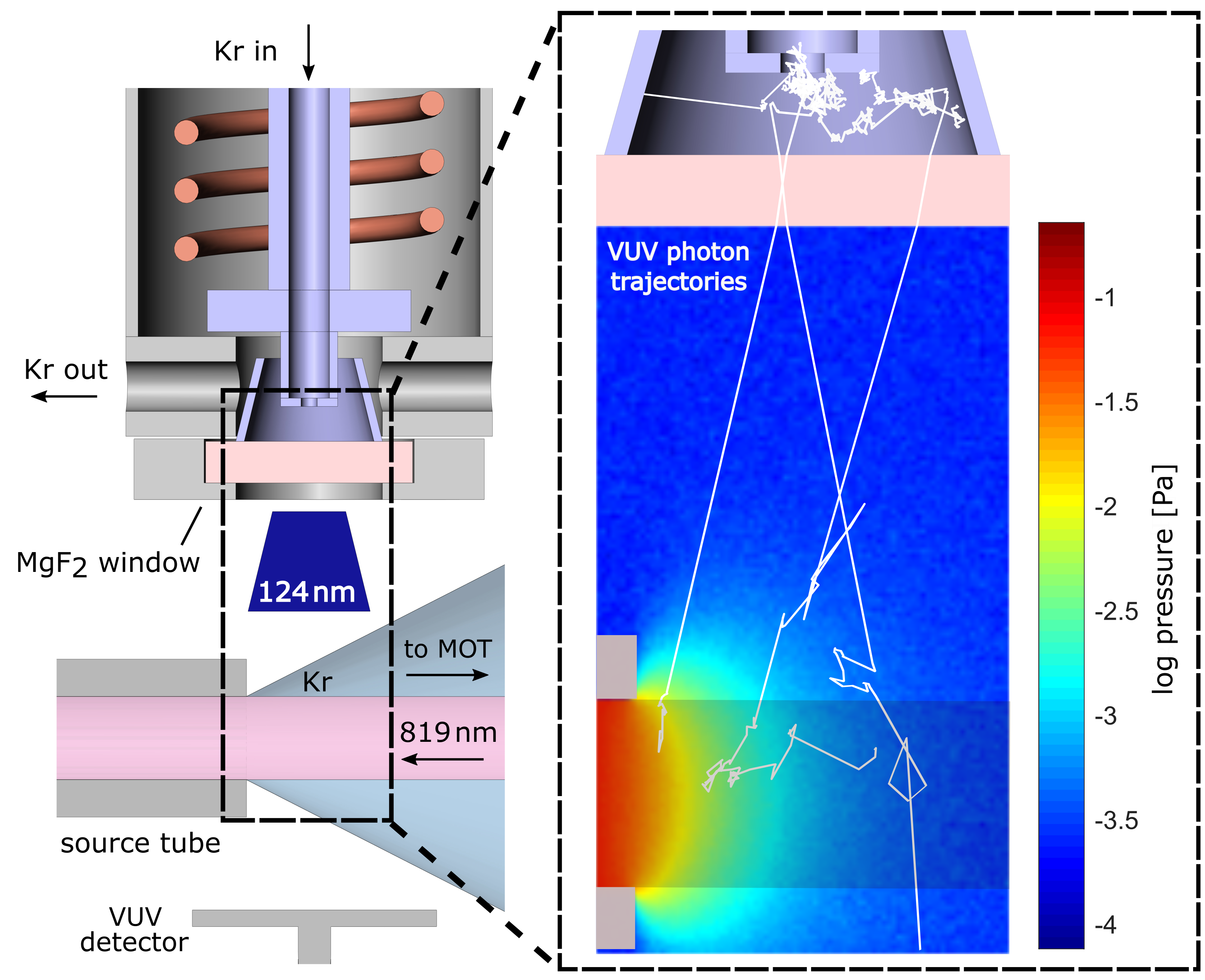}
  \caption {Optical excitation implemented on an ATTA setup for \Kr{81} detection. The \SI{124}{nm} light is generated with a discharge lamp oriented perpendicular to the atomic beam. The \SI{819}{nm} light is directed counter-propagating to the atomic beam. A VUV detector below the lamp measures the VUV intensity. On the right side, a close-up view of the excitation region is shown including the pressure distribution at the outlet of the source tube. The \SI{819}{nm} beam is indicated by the shaded area. Typical \SI{124}{nm}-photon trajectories, generated by a Monte-Carlo simulation, are shown in white.}
  \label{fig:setup}
\end{figure}
\indent Instead of relying on a VUV laser, the \SI{124}{nm} photons can also be generated by a krypton discharge lamp with the commercial krypton in the lamp being strictly separated from the sample krypton in the measurement chamber. The production of metastable krypton using this approach was demonstrated both in a cell \cite{Young2002, Daerr2011} and a beam \cite{Ding2007}. It was also used to demonstrate laser trapping of the abundant isotopes of krypton \cite{Kohler2014}.  However, the lack of a lamp with sufficiently high VUV intensity has so far precluded the implementation of optical excitation in an ATTA system for \Kr{81} detection. A particular difficulty is self-absorption of the \SI{124}{nm} photons in the lamp. According to previous analyses based on Lambert-Beer type calculations, self-absorption leads to a reduced intensity as well as a reversed line profile \cite{Cowan1948, West1976, Weiss1997,Young2002}. With Monte-Carlo simulations of the photon transport in the lamp, we find that the absorbed photons are not lost but that through a multiple scattering process their frequencies are changed. This finding is confirmed by measurements, which agree with simulation results for the VUV output spectrum and the metastable excitation efficiency. The discharge lamp optimized based on these findings has allowed us to realize all-optical Atom Trap Trace Analysis of \Kr{81} at the natural abundance level.\\  \\ \\

\textit{Experimental setup.--- }For trapping and detection of individual \Kr{81} atoms we have developed an ATTA system similar to that described in \cite{Jiang2012, Zhang2020} but customized  to the deployment of optical excitation. The single \Kr{81} atoms are detected via their \SI{811}{nm} fluorescence in the MOT. The loading rate of an abundant isotope is determined by first clearing the MOT with the quenching transition and then measuring the initial linear part of the rising slope of the MOT fluorescence \cite{Cheng2013}. The krypton discharge lamp is installed at the outlet of the source tube as illustrated in Fig.\hspace{0.1cm}\ref{fig:setup}.  The \SI{819}{nm} laser beam is directed along the axis of the atom beam to enable a long interaction region illuminated by the \SI{124}{nm} light from the lamp. The discharge lamp is powered by a RF coil and designed in flow configuration, so that contaminants produced in the discharge are continuously carried away. The VUV photons exit the lamp through a magnesium-fluoride (MgF$_2$) window, which has a transmission of about \SI{70}{\%} at \SI{124}{nm}. A quartz tube (\SI{5}{mm} inner diameter) with a hole at the bottom (\SI{2}{mm}) enables a strong pressure gradient between the upper discharge part and the $\sim$\hspace{0.05cm}\SI{2}{mm} gap right on top of the MgF$_2$ window. This design has proven to be important for maintaining a strong discharge while, close to the window, the number of scattering events are minimized (see following section).\begin{figure}[b]
  \centering
  \noindent  \includegraphics[width=7.5cm]{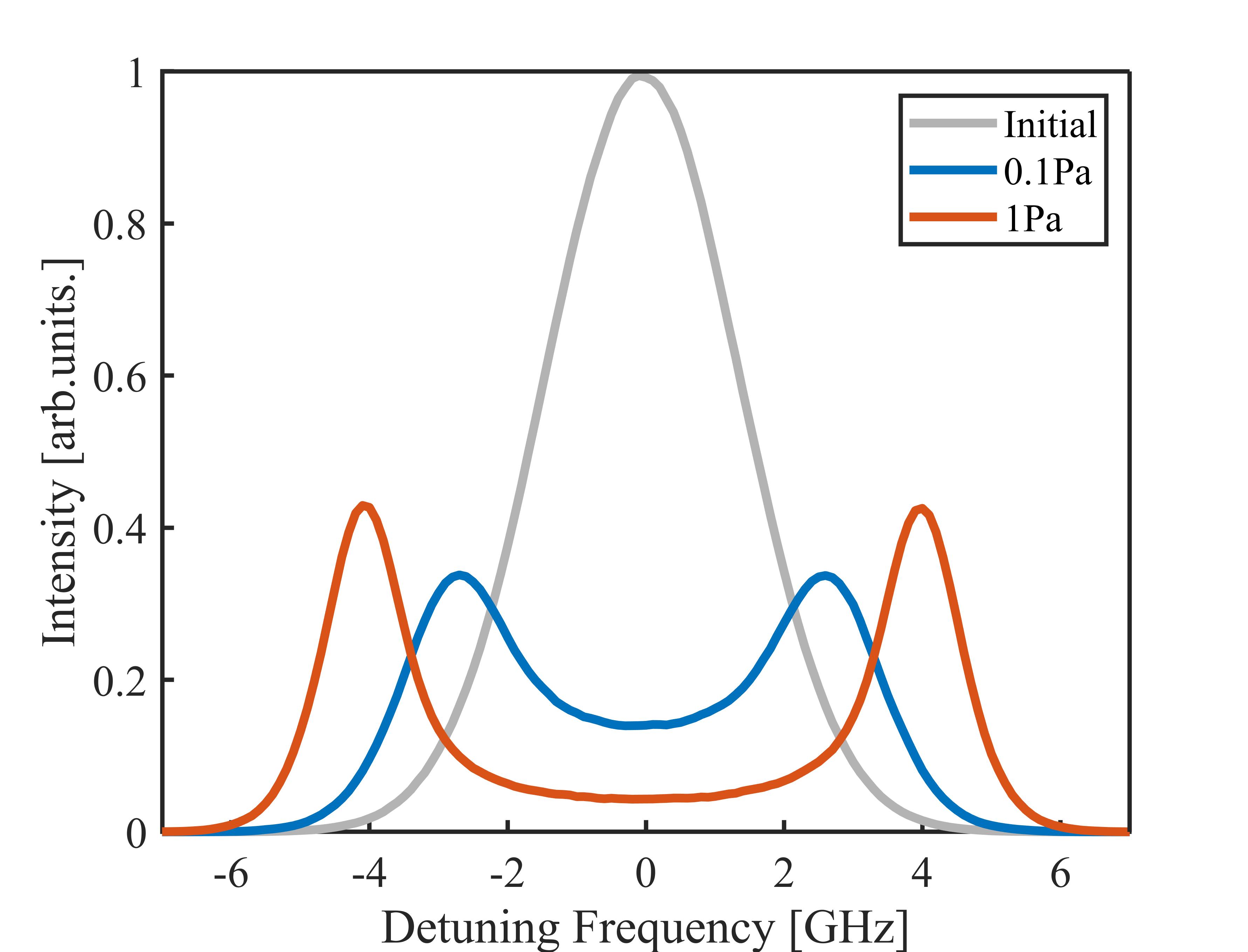}
  \caption {Simulated emission spectra of the lamp at different krypton pressures experimentally relevant to this work. The initial distribution is a Doppler-broadened spectrum at room temperature.}
  \label{fig:detuning}
\end{figure}
\begin{figure*}[t]
  \centering
  \noindent  \includegraphics[width=17cm]{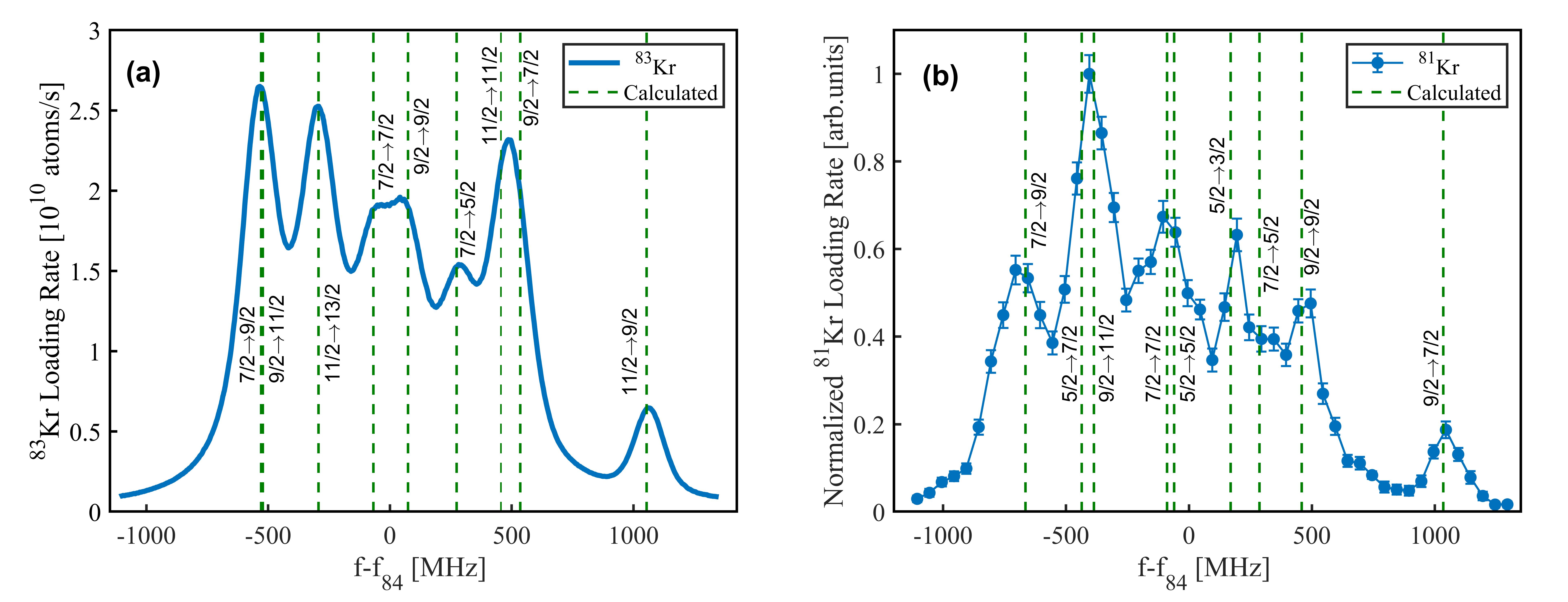}
  \caption { (a) \Kr{83} loading rate vs. frequency of the \SI{819}{nm} laser relative to the \Kr{84} resonance. (b)  \Kr{81} loading rate vs. \SI{819}{nm} frequency. To cancel drifts in the trapping efficiency during the long measurement duration (\SI{3}{days}), the loading rate of the rare \Kr{81} is normalized by the loading rate of the abundant \Kr{83} (which also has hyperfine structure) measured at a fixed laser frequency. The normalized \Kr{81} loading rate is given in arbitrary units. Its error is mainly given by the atom-counting statistics. As \Kr{83} is abundant, its loading rate is measured with \SI{2}{MHz} frequency resolution whereas the single atom loading rate of the rare \Kr{81} is measured in steps of \SI{50}{MHz}. The Doppler shifts due to the longitudinal motion of the atoms are corrected in (a) and (b) to obtain the frequency shifts at rest.}
  \label{fig:Kr83}
\end{figure*} A quartz liner prevents the discharge from impinging on the metal support, which would lead to deposition of contaminants on the MgF$_2$ window. The quartz body as well as the MgF$_2$ window are sealed with resilient aluminum gaskets. The VUV intensity of the lamp is measured via the photoelectric emission from a biased stainless steel plate below the lamp. The measured photon flux on the detector is typically $\sim$\hspace{0.05cm}2$\times$$10^{15}$ photons/s, translating to a photon flux density in forward direction of $\sim$\hspace{0.05cm}3$\times$$10^{16}$ photons/s/sr, comparable to that reported in previous works \cite{Young2002, Okabe1964}. The key development in this work has been on improving the resonance without compromising irradiance.\\

\textit{Simulated frequency spectrum of the lamp.--- }Due to scattering processes in the discharge lamp, the output spectrum is strongly altered from the initial Doppler distribution \cite{Gudimenko2012, Goto2017}. In order to understand the process and its dependencies, we have performed Monte-Carlo simulations of the VUV photon scattering in the lamp and the resulting frequency shifts. Photons are generated with a random frequency according to the Doppler distribution at the outlet of the quartz tube.
 The trajectories of the photons through the lamp are traced until they either escape through the MgF$_2$ window or hit the quartz wall. Typical \SI{124}{nm}-photon trajectories in the lamp (Fig.\hspace{0.1cm}\ref{fig:setup}) show that the photons undergo a large number of scattering events before they obtain a sufficiently large detuning to cross the krypton gas and leave the lamp.
The simulated emission spectra are shown in Fig.\hspace{0.1cm}\ref{fig:detuning}. Interestingly, the total VUV intensity does not change significantly for different lamp pressures. For \SI{0.1}{Pa}, it is almost the same as for \SI{1}{Pa}, amounting to about half of the initial intensity (gray line in Fig.\hspace{0.1cm}\ref{fig:detuning}). This is contrary to the calculation based on Lambert-Beer’s law, which yields a decrease in the VUV intensity by several orders of magnitude as the pressure increases. The simulation shows that at a higher lamp pressure the VUV intensity is not decreasing but that the output spectrum is becoming off-resonant. Consequently, a lamp design that can maintain a high irradiance at a low krypton pressure close to the MgF$_2$ window, as realized in this work, is needed.\\ \indent
The simulation continues to follow the \SI{124}{nm}-photons as they exit the lamp and are scattered in the atomic beam leading to the production of metastable krypton atoms. The pressure distribution at the exit of the source tube is obtained using a test-particle Monte-Carlo simulator for vacuum systems \cite{Kersevan2009}. The resulting pressure distribution is shown in Fig.\hspace{0.1cm}\ref{fig:setup} together with typical \SI{124}{nm}-photon trajectories. The \SI{124}{nm} photons travel relatively undisturbed from the lamp to the exit of the source tube due to the low pressure along the path and the large detuning when leaving the lamp. Upon entering the higher pressure region at the exit of the source tube, the scattering probability increases drastically. After the first scattering event in the higher pressure region, the \SI{124}{nm} photons are likely to become more resonant and will therefore be scattered until combining with an \SI{819}{nm} photon to produce a metastable krypton atom. A similar result has previously been obtained by a Monte-Carlo study in a krypton cell \cite{Young2002}, including effects of photon scattering and energy redistribution.\\

\begin{figure}[t]
  \flushleft
  \noindent  \includegraphics[width=8.5cm]{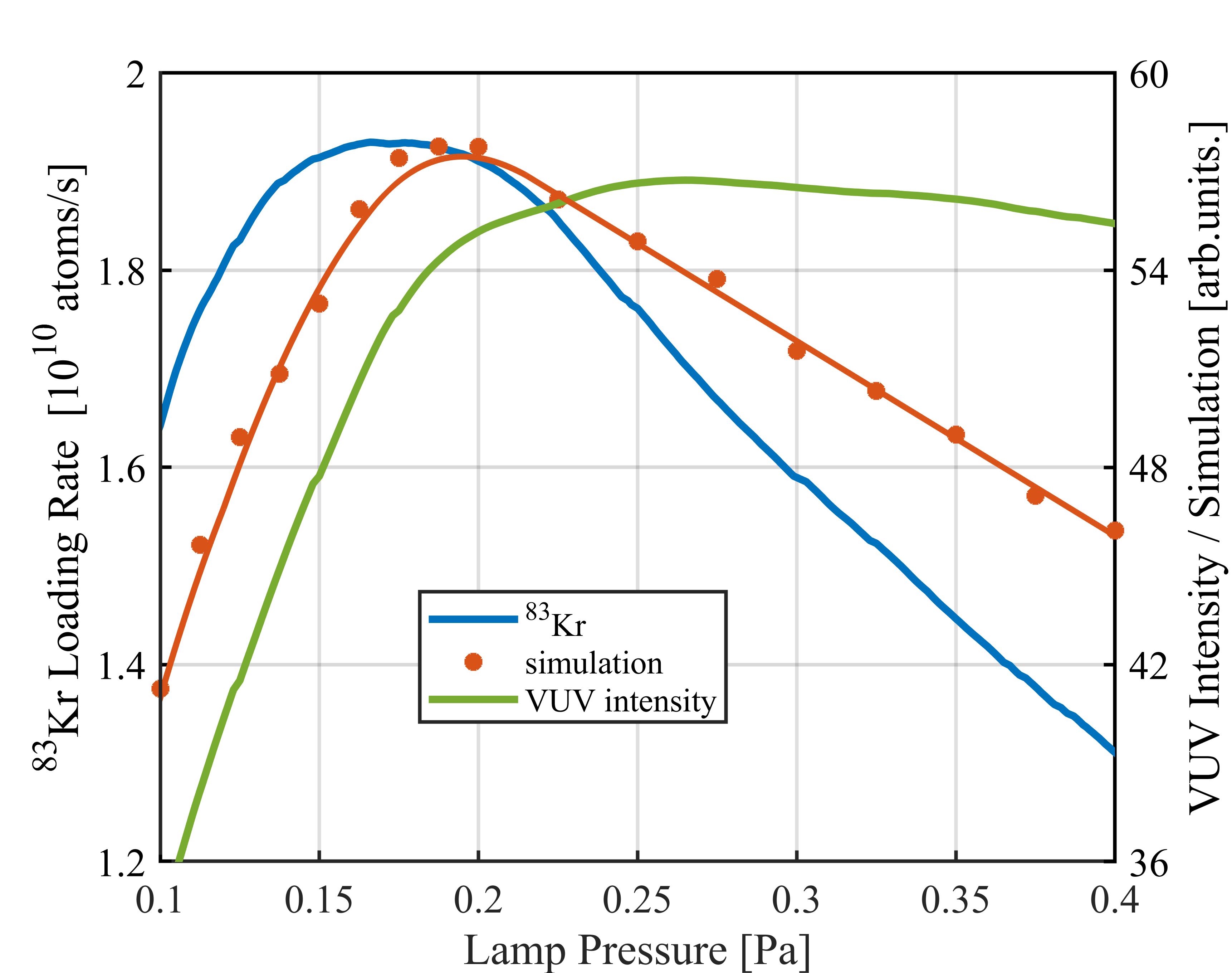}
  \caption { \Kr{83} loading rate, VUV intensity and simulated number of produced metastable krypton atoms vs. lamp pressure. The orange line is a guide-to-the-eye for the simulated points (also orange).}
  \label{fig:plamp}
\end{figure}

\textit{Results and discussion.--- }The MOT loading rate of the abundant \Kr{83} (\SI{11.5}{\%} isotopic abundance) is measured while the frequency of the \SI{819}{nm} laser is scanned across the resonances of the hyperfine manifolds of the $1\text s_4$\hspace{0.05cm}$-$\hspace{0.05cm}$2\text p_6$ transition (Fig.\hspace{0.1cm}\ref{fig:Kr83}a). The highest loading rate occurs at the overlap of the transitions $ F=7/2\rightarrow9/2 $ and $ F=9/2\rightarrow11/2 $. The $ F=11/2\rightarrow13/2 $ transition is similarly high as expected from the transition strength \cite{Axner2004}. In order to address not only one but all three hyperfine levels, we generate sidebands on the \SI{819}{nm} light with an electro-optical modulator. When the carrier is resonant with the $ F=11/2\rightarrow13/2 $ transition, we obtain an increase in the loading rate of \SI{35}{\%} with a modulation frequency of \SI{240}{MHz}, i.e. when the sidebands are resonant with the two neighboring transitions. The intense and resonant output of the discharge lamp allows us to trap and detect the extremely rare isotope \Kr{81} (9.3$\times$$10^{-13}$ isotopic abundance \cite{Zappala2020}) using optical excitation instead of discharge excitation. Fig.\hspace{0.1cm}\ref{fig:Kr83}b shows the normalized \Kr{81} loading rate vs. frequency of the \SI{819}{nm} laser, which confirms the transition frequencies calculated in \cite{Zhang2020}.\\
\indent Fig.\hspace{0.1cm}\ref{fig:plamp} shows the dependence of the \Kr{83} loading rate and the VUV intensity on the krypton pressure in the lamp. The lamp pressure (in the region on top of the window) is calculated according to the pressure measured downstream and the relevant flow conductances. At a lamp pressure of \SI{0.2}{Pa}, the \Kr{83} loading rate is $\sim$\hspace{0.05cm}\SI{50}{\%} higher than at \SI{0.4}{Pa}, whereas the VUV intensity is almost the same.
This behavior supports the result of the simulations in Fig.\hspace{0.1cm}\ref{fig:detuning} (note the similar pressure range) that the photons are not lost at a higher lamp pressure but only become less resonant. When the lamp pressure drops below \SI{0.2}{Pa}, the \Kr{83} loading rate decreases sharply in accordance with the strongly decreasing VUV intensity. Based on the measured VUV intensity as a function of lamp pressure (Fig.\hspace{0.1cm}\ref{fig:plamp}) and the simulation of the VUV photon trajectories described above, we obtain the simulated number of produced metastables as a function of lamp pressure (Fig.\hspace{0.1cm}\ref{fig:plamp}). As the \Kr{83} loading rate is proportional to the number of metastable atoms the two curves (blue and red) should scale together, which is indeed the case. The small difference between the optimum pressures may be due to that the lamp pressure close to the window region is systematically higher than the calculated value. The general agreement of the simulated and measured dependence on the lamp pressure is a further confirmation of the photon transport process described above (see also Supplemental Material). \\ 
\begin{table}[t!]
	\caption{Loading rates of the different krypton isotopes for an atmospheric krypton sample using all-optical ATTA.\\}
	\centering
	\def\arraystretch{1.2} 
	\begin{tabular*}{\hsize}{@{}@{\extracolsep{\fill}}ccc}
		\hline\hline
		Isotope                                                           & Isotopic abundance     & Loading rate                                             \\ \hline
		\begin{tabular}{@{}l@{}}\Kr{84} \end{tabular} & \SI{57.0}{\%}     & {4.5}$\times$\SI{e11}{atoms/s}   \\ 
		\begin{tabular}{@{}l@{}}\Kr{83} \end{tabular} & \SI{11.5}{\%}     &{7.2}$\times$\SI{e10}{atoms/s}    \\ 
		\Kr{85}                                       &  {1.5}$\times$\SI{e-11}  & {2.6}$\times$\SI{e4}{atoms/h}             \\ 
	    \Kr{81}                                       & {9.3}$\times$\SI{e-13}       & {1.8}$\times$\SI{e3}{atoms/h}   \\ \hline\hline
	\end{tabular*}
\label{tab:counting_rate}
\end{table}
\indent The loading rates of the odd krypton isotopes, measured using optical excitation under optimum conditions, are shown in Table \ref{tab:counting_rate}. To assess the influence of hyperfine structure on the excitation efficiency, the loading rate of the abundant \Kr{84} (no hyperfine structure) is also measured. The loading rate ratio between \Kr{84} and \Kr{83} is a little higher than their isotopic abundance ratio indicating that the sideband coverage for the odd \Kr{83} is incomplete. The \Kr{83} loading rate in Fig.\ref{fig:Kr83} is lower than that in Table \ref{tab:counting_rate} because no sidebands on the \SI{819}{nm} light were added and the transmission of the MgF$_2$ window had already decreased due to prior usage. The achieved single atom loading rates for the rare \Kr{81} and \Kr{85} ($t_{1/2}=\SI{11}{a}$) are comparable to state-of-the-art ATTA systems based on discharge excitation \cite{Jiang2019}.\\
\indent The VUV intensity of the lamp presented here drops to about \SI{50}{\%} in the first 20 hours. In the following 100 hours the output decreases linearly to about \SI{30}{\%} of the initial intensity. The decrease might be caused by part of the discharge still reaching metal parts. It may thus be mitigated by making the lamp entirely with glass and the MgF$_2$ window \cite{Freeman1978, Nevyazhskaya2012}. Another contribution to the decreasing transmission of the MgF$_2$ window may be the formation of color centers in the MgF$_2$ crystal induced by the VUV radiation \cite{Sibley1968, Zhukova2002}. However, the achieved lifetime of the present lamp is already sufficient for operational use.\\ 
\indent As described in the introduction, the conventional discharge-excitation leads to sample loss and cross-sample contamination due to sputtering processes. Over a typical \Kr{81} measurement of 4 hours, about \SI{70}{\%} of a \SI{1}{\mu L} krypton sample is lost and $\sim$\hspace{0.05cm}\SI{30}{nL/h} of contaminant krypton is sputtered out of the system. We have measured these characteristics in our all-optical setup and find that both limitations are overcome. When filling in a \SI{1}{\mu L} krypton sample, the loss after a 4 hour measurement is below the detection limit of \SI{5}{\%}. The krypton contamination rate is only \SI{0.1}{nL/h}, two orders of magnitude lower than that of the discharge excitation. Within measurement uncertainty the contamination rate is the same with and without lamp. \\

\textit{Conclusion and outlook.--- }With the all-optical approach we achieve a \Kr{81} loading rate similar to state-of-the-art systems based on discharge excitation while sample loss and cross-sample contamination are negligible. Consequently, the measurement time can be extended from a few hours to a few days, gaining counting statistics accordingly. The sample size requirement of \Kr{81} dating can thus be brought to its fundamental limit below \SI{1}{kg}, with the precision given by the number of \Kr{81} atoms contained in a water or ice sample. These advances enable new applications for \Kr{81} dating in the earth sciences, particularly for polar ice cores where high precision is needed and the sample size is limited.  Due to the low background, all-optical ATTA can also be used to analyze sub-ppt levels of krypton in the purified xenon for the next generation of dark matter detectors \cite{Lindemann2014, Aprile2018, Aprile2019}. The excitation technique realized here can readily be applied in other metastable krypton or xenon experiments. It is more demanding for metastable argon since the required VUV wavelength (\SI{107}{nm}) is shorter, requiring a lithium-fluoride window for sufficient transmission.  \\
\indent It is straightforward to further increase the optical excitation efficiency by adding more lamps. This will further reduce the measurement time for \Kr{81} analysis and opens a new pathway for high number density metastable noble gas experiments.\\


{\textit Acknowledgements. }This work is funded by the National Key Research and Development Program of China (2016YFA0302200), National Natural Science Foundation of China (41727901, 41861224007, 41961144027), Anhui Initiative in Quantum Information Technologies (AHY110000). We thank Xing-Yu Li for his help on part of the lamp measurements. \\ \\
\indent \textit{An edited version of this paper was published by APS \href{https://link.aps.org/doi/10.1103/PhysRevLett.127.023201}{Physical Review Letters \textbf{127}, 023201 (2021)}. Copyright 2021 American Physical Society.}

	
\bibliographystyle{apsrev4-2}

\end{document}


\title{\Large Supplemental material \\\vspace{0.7cm} {\large Optical excitation and trapping of \Kr{81} }}
\author{J. S. Wang}

\author{F. Ritterbusch}
\email{florian@ustc.edu.cn}

\author{X.-Z. Dong}
\author{C.-Y. Gao}
\author{W. Jiang}
\author{S.-Y. Liu}

\author{Z.-T. Lu}
\email{ztlu@ustc.edu.cn}

\author{G.-M. Yang}
\author{Z.-Y. Zhang}

\affiliation{\medskip Hefei National Laboratory for Physical Sciences at the Microscale, CAS Center for Excellence in Quantum Information and Quantum Physics, University of Science and Technology of China, 96 Jinzhai Road, Hefei 230026, China}

\maketitle

\renewcommand{\thefigure}{S\arabic{figure}}

\setcounter{figure}{0}

\begin{figure}[h!]
  \centering
  \noindent  \includegraphics[width=9cm]{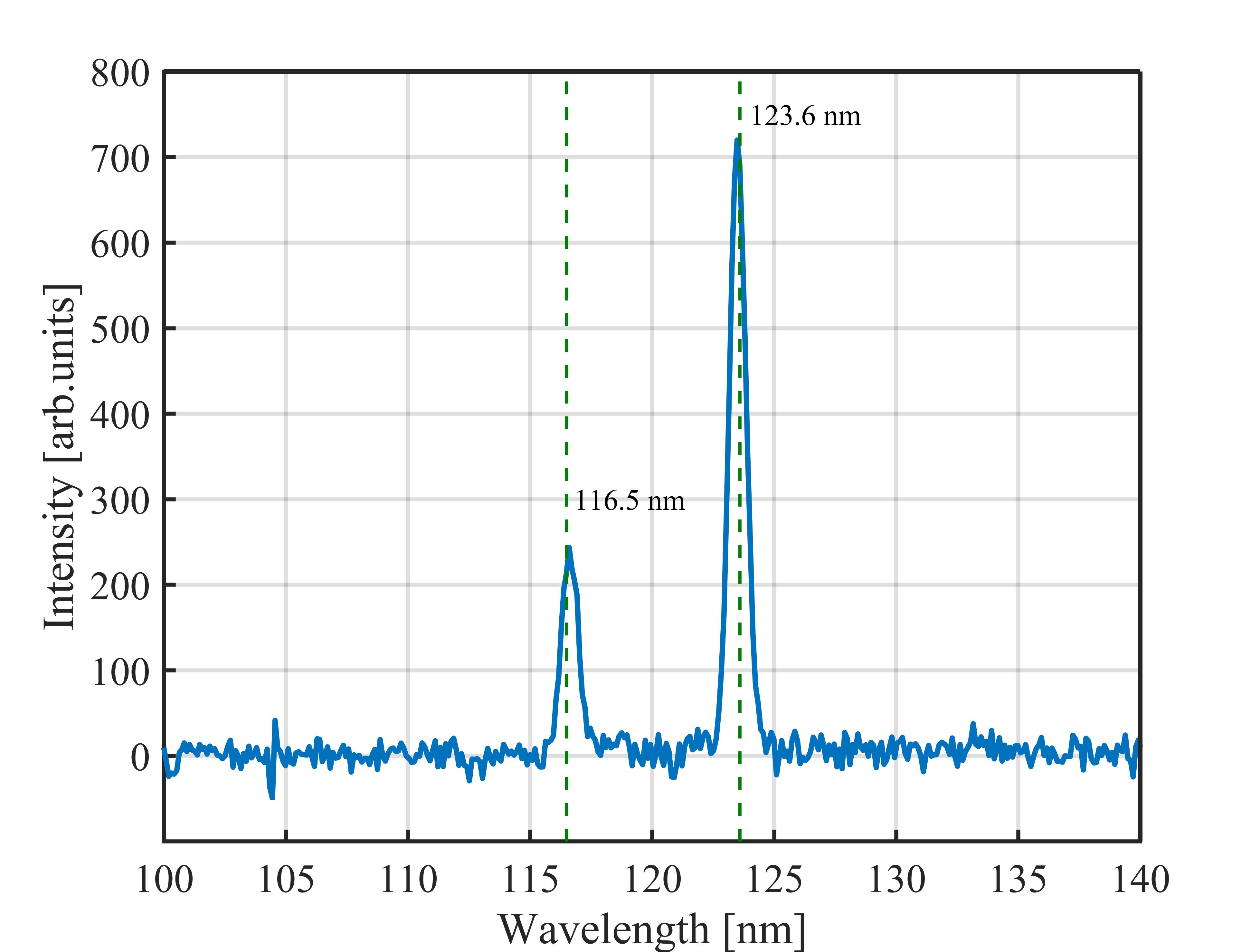}
  \caption {Output spectrum of the lamp measured with a monochromator.   }
  \label{fig:output_spectrum}
\end{figure}

The output spectrum of the lamp is measured with a monochromator (resolution $\sim$\hspace{0.05cm}\SI{1}{nm}) using a biased stainless steel plate as a detector (Fig. \ref{fig:output_spectrum}). With a work function of \SI{4.4}{eV}, the detector is only sensitive to photons with a wavelength shorter than \SI{280}{nm}, a range within which krypton only has the \SI{116}{nm} and \SI{124}{nm} lines originating from the $1s_2$ and $1s_4$ levels, respectively. From the lamp output spectrum in Fig. \ref{fig:output_spectrum}, it can be seen that the \SI{124}{nm} line is $\sim$\hspace{0.05cm}3 times stronger than the \SI{116}{nm} line. Taking into account the wavelength dependent transmission of the MgF$_2$ window, this indicates that the populations in level $1s_4$ and $1s_2$ are similar.

\begin{figure}[!h]
  \noindent  \includegraphics[width=8cm]{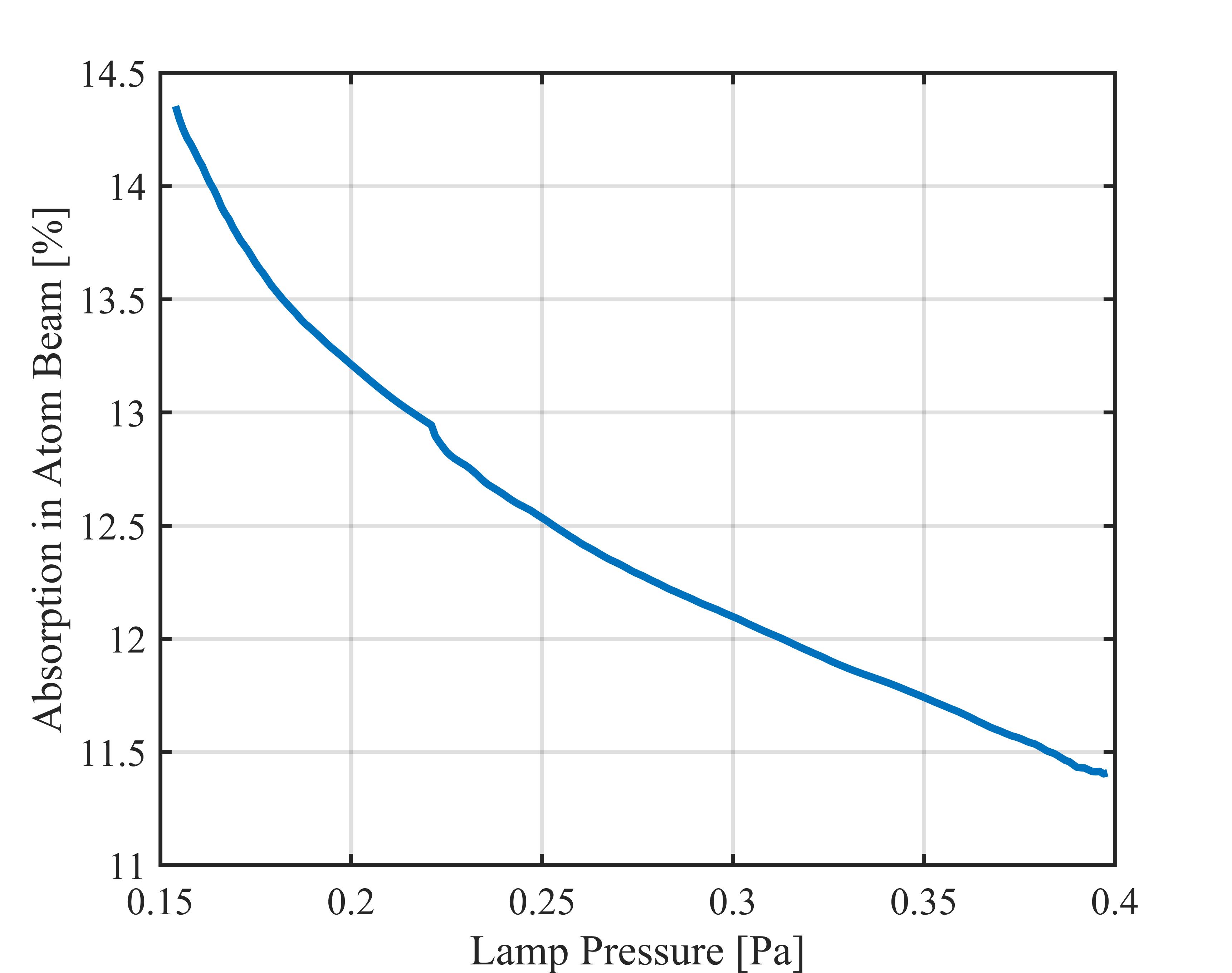}
  \caption {Absorption of the VUV light in the atom beam vs. lamp pressure.  }
  \label{fig:absorption}
\end{figure}

An essential design criteria for the lamp in this work has been to improve its resonance without compromising irradiance. According to the simulation shown in Fig. 3, this can be reached if the design allows a low krypton pressure in the lamp while maintaining a high irradiance. 
The lamp developed in this work shown in Fig. 2 realizes this condition due to the steep pressure gradient between the high pressure discharge region upstream and the low pressure region close to the MgF$_2$ window. Fig. 5 confirms the conclusion that the VUV photons are not lost due to the scattering processes but that their frequencies are redistributed. This means that the lamp pressure plays a critical role in determining the resonance of the lamp output. The results shown in Fig. \ref{fig:absorption} provide a further confirmation of this finding. The VUV output power is measured by the VUV detector (Fig. 2) at different lamp pressures with and without the krypton atomic beam. The resulting absorption of the VUV light in the atom beam as a function of lamp pressure is plotted in Fig. \ref{fig:absorption}. It is apparent that the absorption increases with decreasing lamp pressure, i.e. the lamp light becomes more resonant as predicted by the simulation results shown in Fig. 3.   
